\documentclass{article}
\usepackage{xspace}
\usepackage{bsymb}
\usepackage{natbib}
\newcommand{\eb}{\textsc{Event-B}\xspace}
\newcommand{\cb}{\textsc{B-method}\xspace}
\newcommand{\alloy}{\textsc{Alloy}\xspace}
\newcommand{\pcite}[1]{(\cite{#1})}
\title{Model Checking Event-B by Encoding into Alloy\footnote{This work is
    partially supported by EPSRC grant EP/E012973/1, and by EU grants
    IST/033709 and ICT/217069.}}
\author{Paulo J. Matos $\qquad$ Jo\~ao Marques-Silva \\
Electronics and Computer Science, University of Southampton\\
\texttt{\{pocm, jpms\}@ecs.soton.ac.uk}}
\begin{document}
\maketitle
\begin{abstract}
As systems become ever more complex, verification becomes more main
stream. \eb and \alloy are two formal specification languages based on
fairly different methodologies. While \eb uses theorem provers to
prove that invariants hold for a given specification, \alloy uses a
SAT-based model finder. In some settings, \eb invariants may not be
proved automatically, and so the often difficult step of interactive
proof is required. One solution for this problem is to validate
invariants with model checking. This work studies the encoding of \eb
machines and contexts to \alloy in order to perform temporal model
checking with \alloy's SAT-based engine.
\end{abstract}
\section{Introduction}
\label{sec:intro}
%- Intro
Current day systems are ever more detailed and complex leading to the
necessity of developing models that abstract unimportant
implementation details while emphasizing their structure. These
models are developed in order to be easily verified either by theorem
provers or model checkers.

%- EventB
\eb is a language based on \cb\pcite{abrial-book96}, and supported by
the open tool RODIN~\footnote{The RODIN toolkit was developed under
  research project IST 511599.}.
Part of the language is developed visually and there
is a syntax for predicates and
expressions\pcite{rodind7-techrep05}. The RODIN tool is shipped with
theorem provers which allow the user to prove the model
invariants either automatically or interactively.

%- Alloy
\alloy~\pcite{jackson-tosem02} is a structural modeling language based
in first order logic. It is a textual language that consists of
signatures, which introduce flat relations, functions, predicates and
assertions that deal with the relations. By specifying predicates and
assertions it is possible to perform model finding or model checking.
The tool uses KodKod~\pcite{torlak-tacas07}, a model finder, to convert
the model within given bounds to SAT and return an instance of the
model or a counter-example. 

%- Previous
In previous work~\pcite{butler-zb02} \cb was combined with \alloy.
since then \alloy was considerably improved and extended by, not only,
adding new constructs but also by making the verification much more
efficient. The restrictions described in the work either do not exist
anymore or they can be overcome.

%- Motivational content
Until recently it was only possible to perform temporal model checking
in an \eb model by using a two step process: converting the
model to \cb and then using ProB\pcite{butler-fme03}. However,
the solution was not straightforward as the conversion to \cb was not
always fully automated. More recently, a prototype ProB
plugin\pcite{leuschel-afadl07} for the RODIN tool has been developed
that provides an alternative solution for model checking \eb models
using the ProB model checker.
Nevertheless, encoding \eb to \alloy allows building on top of the
\alloy model finding engine therefore benefiting from all of its 
optimizations. As mentioned elsewhere~\pcite{leuschel-afadl07},
encoding \eb is not straightforward but this work shows it is
possible. This work has been published in~\pcite{pmatos-abz08}.
\section{Encoding}
\label{sec:encoding}
%- Encoding
This section summarizes the process of encoding an \eb model into
\alloy. Due to space constraints the presentation will 
be informal and far from complete, which means that there
are \eb operators for which an encoding has been developed but which
are not described. A full example, based on processes and mutexes,
including the actual encoding and additional comments will be shown in
section~\ref{sec:example}.
For a detailed description of the structure and rules of \eb models
the reader is referred to~\pcite{rodind7-techrep05}. 

%- Introduction
There are three aspects to the encoding: encoding of the model
structures, expressions, and predicates (which are straightforward due
given the existence of the logical operators in both languages).

%- Structure
The execution model needs to be emulated by the final \alloy 
model. To this end all of the encoded models define a
``State'' signature which is ordered by using the ordering module and
with as many fields are there are variables. The variable types are
extracted from the set of \eb invariants and encoded into signatures
which become the type of the respective fields. A fact defines the
initial state which is encoded from the ``Initialisation'' event. Each
event is encoded in a predicate with two arguments: the current state
and the next state and it evaluates to true whenever the next state
reflects the triggering of an event from the current state. A final
fact asserts that at every state one of the events needs to be
triggered. 

A carrier set is encoded as a signature with no fields and an
enumerated set is encoded as signature, one per enumeration, with no
fields that extend a base signature that represents the type of the
enumerated set.  One special enumerated set ``Events''
 has one enumeration per event, plus an ``Undef'' enumeration for
the initial state which is defined to be triggered by an ``Undef''
event. ``Ev'' is the type of a special field in the ``State''
signature. If ``Ev'' is $x$ in state $s'$, this means that it was
the triggering of $x$ that caused the transition from $s$ to
$s'$. Although this information is not necessary for the model
checking itself, it makes the state trace of a failed invariant much
more readable.

%- Expressions
Expressions are the hardest part to encode. There is not only a myriad
of complex expressions in \eb but given that \alloy uses only flat
relations, some \eb expressions that introduce relations with nested
sets generate many (and potentially large) \alloy expressions. Some
expressions are straightforward like the domain, range, domain and
range restriction and their subtraction counterparts, since they are
either already defined in modules shipped as part of the current
\alloy distribution or they are very easily defined as small
functions. Operators like $\prjone$, $\prjtwo$ and $\id$ need to be
defined as \alloy functions in order to be used. All arithmetic
operators except power are defined but still, power can be defined
explicitly during encoding-time depending on the defined bit width
passed on to the \alloy engine for checking. For example, $a^b$ could
be defined as $a = 0 \limp 1 \textrm{ else } b = 0 \limp 0 \textrm{
  else } b = 1 \limp a \textrm{ else } b = 2 \limp a.mult[a] \cdots$.
Function expressions can be encoded as relations and then facts can be
added to the model as to assure the semantics is preserved. So, to
encode $(A \tfun B) \rel C$, a signature with a relation from $A$ to
$B$ would be defined followed by a fact asserting the relation to be
a total function and then yet another signature is defined with a relation from
the previously defined signature to $C$. Although function nesting
requires a new signature, it seems this is the best general solution
given that \alloy only works with flat relations.
\section{Example}
\label{sec:example}
%- Encoding Example
This section presents an example of the encoding. Although the example
is simple, it is enough to have an idea of how it works. The example
is based on the Alloy model \texttt{dijkstra.als} distributed with
Alloy~4.1.

The \eb specification is shown in table~\ref{tb:eventb}. It introduces
two carrier sets, the processes set and the mutexes set, and two
relations between processes and mutexes. The pair $\{p \mapsto m\}$ is
in the relation $Holds$ iff the process $p$ holds the mutex $m$ and it is in
the relation $Waits$ if $p$ is waiting for $m$ to be released in order
to hold it. There are three events which control the holding and
release of mutexes by processes. The ``HoldOnMutex'' event is triggered
for a process $p$ and a mutex $m$ if $p$ is not waiting for
any mutex and it does not already hold $m$. Once it is
triggered $p$ holds $m$ by adding the pair $\{p \mapsto m\}$ to the
$Holds$ relation. The ``WaitOnMutex'' event is triggered when process $p$ has to wait to hold $m$
because there is already some other process holding it. The
``ReleaseMutex'' is the counterpart to ``HoldOnMutex'' and allows a process to release a mutex it no longer
needs. This very simplistic model allows us to exercise the encoding
presented in the paper.
\begin{table}
\caption{Mutexes \eb Specification}\label{tb:eventb}
\begin{center}
\begin{tabular}{p{0.5cm}l|p{0.5cm}l}
\hline
\multicolumn{2}{l}{{\bfseries Carrier Sets}}& \multicolumn{2}{|l}{{\bfseries Event} HoldOnMutex} \\
                         & $Process$ & \multicolumn{2}{l}{$\;${\bfseries \emph{Guards}}}                   \\ 
                         & $Mutex$   &                                & $p \in Process$         \\
\multicolumn{2}{l|}{{\bfseries Variables}} &                           & $m \in Mutex$           \\
                         & $Holds$   &                                & $p \not\in \dom(Waits)$ \\
                         & $Waits$   &                                & $m \not\in \ran(Holds)$ \\
\multicolumn{2}{l|}{{\bfseries Initialisation}} & \multicolumn{2}{l}{$\;${\bfseries \emph{Actions}}}            \\
                         & $Holds \bcmeq \emptyset$ &           & $Holds \bcmeq Holds \bunion \{ p \mapsto m \} $ \\
                         & $Waits \bcmeq \emptyset$ & \multicolumn{2}{|l}{{\bfseries Event} WaitOnMutex} \\
\multicolumn{2}{l|}{{\bfseries Event} ReleaseMutex} & \multicolumn{2}{l}{$\;${\bfseries \emph{Guards}}} \\
\multicolumn{2}{l|}{$\;${\bfseries \emph{Guards}}}             &                      & $p \in Process$         \\
                         & $p \in Process$          &                 & $m \in Mutex$           \\
                         & $m \in Mutex$           &                 & $p \not\in \dom(Waits)$ \\
                         & $p \not\in \dom(Waits)$  &                 & $m \in \ran(\{p\} \domsub Holds)$ \\
                         & $m \in \ran(\{p\} \domres Holds)$ & \multicolumn{2}{|l}{$\;${\bfseries \emph{Actions}}}   \\
\multicolumn{2}{l|}{$\;${\bfseries \emph{Actions}}}             &                 & $Waits \bcmeq Waits \bunion \{p \mapsto m\}$ \\
                         & $Holds \bcmeq Holds \setminus \{p \mapsto m\}$ & \multicolumn{2}{l}{{\bfseries Invariants}}\\
                         &   &                 & $Holds \in Process \rel Mutex$ \\
                         &  &        & $Waits \in Process \rel Mutex$ \\
                         &            &    & $\dom(Waits) \neq Process$ \\
\hline
\end{tabular}
\end{center}
\end{table}
The carrier sets and the enumerated set for event types are
immediately encoded into several signatures as seen in table~\ref{tb:alloyinitsigs}.
\begin{table}
\caption{Carrier Set and Enumerated Set Encodings}\label{tb:alloyinitsigs}
\begin{center}
\begin{tabular}{l}
\hline
sig $Process \{\}$ \\
sig $Mutex \{\}$ \\
abstract sig $Events \{\}$ \\
one sig $Undef$, $HoldOnMutexE$, $WaitOnMutexE$, \\ 
$\quad\quad\quad ReleaseMutexE$ extends $Events \{\}$ \\
\hline
\end{tabular}
\end{center}
\end{table}
The state signature along with the types of the $Holds$ and $Waits$
relation, which are extracted from the invariants is shown in
table~\ref{tb:statesig}.
\begin{table}
\caption{State Signature and Types}\label{tb:statesig}
\begin{center}
\begin{tabular}{p{4cm}|p{4cm}|p{3cm}}
\hline
sig $HoldsRel \{$              &sig $WaitsRel \{$            &sig $State \{$ \\
$\;rel : Process \tfun Mutex$  &$\;rel : Process \tfun Mutex$&$\;Holds : HoldsRel,$\\
$\}$                           &$\}$                         &$\;Waits : WaitsRel,$\\
                               &                             &$\;Ev : Events$\\
                               &                             &$\}$ \\
\hline
\end{tabular}
\end{center}
\end{table}
Although it would be possible to define the relations $Holds$ as
$Holds : Process \tfun Mutex$ and analogously for $Waits$ we didn't do
that in this case because the encoding of \eb expressions as $((A
\pfun B) \tsur C)$ is done recursively on the structure of the
expression generating $n+1$ signatures for $n$ operators. 

The initialisation event is encoded as a fact that sets the initial
state $s0$, and is shown on the left side of table~\ref{tb:eventinittrigger}.
\begin{table}
\caption{Event $Initalisation$}\label{tb:eventinittrigger}
\begin{center}
\begin{tabular}{l|l}
\hline 
$\textrm{fact } Initial \{$                           & fact $EventTrigger \{$                   \\
$\;\textrm{let } s0 = ord/first \{$                   & $\;$all $s : State - ord/last \{$        \\
$\;\;s0.Holds.rel = \textrm{none} \tfun \textrm{none}$& $\;\;\textrm{let } s' = ord/next[s] \{$  \\
$\;\;s0.Waits.rel = \textrm{none} \tfun \textrm{none}$& $\;\;\;HoldOnMutex[s,s'] \textrm{ or } $ \\
$\;\;s0.Ev = \textrm{Undef}$                          & $\;\;\;\;WaitOnMutex[s,s'] \textrm{ or }$\\
$\;\}$                                                & $\;\;\;\;ReleaseMutex[s,s']$             \\
$\}$                                                  & $\;\;\}$                                 \\
                                                      & $\;\}$                                   \\
                                                      & $\}$                                     \\
\hline
\end{tabular}
\end{center}
\end{table}
This encoding just defines the variable $s0$ as the first element of the
state ordered set and then sets each field of the state with its
initial value. On the right side the trigger is a fact that forces for
each state transition one of the events to trigger. 

The events are encoded by translating each of the guards and actions
into \alloy expressions. Table~\ref{tb:eventwaitonmutex} presents the
encoding for the $WaitOnMutex$ event. Most of the encoding is
straightforward, except for the case of the domain subtraction
operator that is encoded into \alloy using its definition based on the
domain restriction operator. Note that in \alloy, whenever the actions
do now span all of the state variables, it is necessary to state that
they are equal to the previous state. 
\begin{table}
\caption{Event $WaitOnMutex$}\label{tb:eventwaitonmutex}
\begin{center}
\begin{tabular}{l}
\hline
$\textrm{pred } WaitOnMutex [s, s' : State] \{$                  \\
$\;\textrm{some } p : Process, m : Mutex \{$                   \\
$\;\;$// Guards                                        \\
$\;\;!(p \textrm{ in dom }[s.Waits.rel])$              \\
$\;\;m \textrm{ in ran}[(\textrm{dom}[s.Holds.rel] - \{p\}) <: s.Holds.rel]$\\
$\;\;$// Action                                        \\
$\;\;s'.Waits.rel = s.Waits.rel + \{p \tfun m\}$       \\
$\;\;s'.Holds = s.Holds$                               \\
$\;\;s'.Ev = WaitOnMutexE$                             \\
$\;\}$                                                 \\
$\}$                                                   \\
\hline
\end{tabular}
\end{center}
\end{table}

The $HoldOnMutex$ events and its counterpart $ReleaseMutex$ are shown
in table~\ref{tb:eventholdonmutex} on the left and right side
respectively. The events are similar and their encoding just
transforms the \eb operators into those recognized by \alloy.
\begin{table}
\caption{Event $HoldOnMutex$ and $ReleaseMutex$}\label{tb:eventholdonmutex}
\begin{center}
\begin{tabular}{l|l}
\hline
$\textrm{pred } HoldOnMutex [s, s' : State] \{$  & $\textrm{pred }ReleaseMutex[s, s' : State] \{$  \\
$\;\textrm{some } p : Process, m : Mutex \{$     & $\;\textrm{some } p : Process, m : Mutex \{$    \\
$\;\;$// Guards                                  & $\;\;$// Guards                                 \\
$\;\;!(p \textrm{ in dom}[s.Waits.rel]) $        & $\;\;!(p \textrm{ in dom}[s.Waits.rel])$        \\
$\;\;!(m \textrm{ in ran}[s.Holds.rel])$         & $\;\;m \textrm{ in ran}[\{p\} <: s.Holds.rel]$  \\
$\;\;$// Action                                  & $\;\;$// Action                                 \\
$\;\;s'.Holds.rel = s.Holds.rel + \{p \tfun m\}$ & $\;\;s'.Holds.rel = s.Holds.rel - \{p \tfun m\}$\\
$\;\;s'.Waits = s.Waits$                         & $\;\;s'.Waits = s.Waits$                        \\
$\;\;s'.Ev = HoldOnMutexE$                       & $\;\;s'.Ev = ReleaseMutexE$                     \\
$\;\}$                                           & $\;\}$                                          \\
$\}$                                             & $\}$                                            \\
\hline
\end{tabular}
\end{center}
\end{table}
Given a syntactically correct \alloy module header, it is possible to
verify the \alloy model by encoding the invariant as an assertion:
\[
\textrm{assert }NoDeadlock \{\textrm{all }s : State\;|\;!(\textrm{dom}[s.Waits.rel] = Process)\}
\]
and using the \textrm{check} command to verify it.
The following check finds a deadlock within 6 states, 2
processes and 2 mutexes. This translates analogously into \eb. There
is an invariant failure within 6 event transitions for 2 processes and
2 mutexes.
\begin{eqnarray*}
\textrm{check }NoDeadlock \textrm{ for exactly } 6\;State,\textrm{
  exactly }2\;Process,\textrm{ exactly }2\;Mutex, \\
\qquad\qquad\qquad\qquad\qquad\textrm{ exactly }6\;HoldsRel,\textrm{ exactly }6\;WaitsRel
\end{eqnarray*}

The second line of the \textrm{check} command specifies necessary bounds
based on the number of states \alloy should
check. The number of such auxiliary signatures is always equal to the
number of states. 
\section{Conclusion and Future Work}
\label{sec:conclusion}
%- Conclusion
The focus of our work is to allow the users of the \eb
language to use the years of work and expertise in the development of the
\alloy tools. This paper summarizes an encoding of \eb into
\alloy. The resulting \alloy model can serve to find counterexamples
to false invariants and translate them back to \eb.
%- Future Work
Future work entails the automatic generation of the encoding and its
integration with the RODIN tool. The tool to be developed can then be
extended to use other backends besides \alloy.
%- Constraint Based Checking
Although this work focuses on model checking, there are cases where it
is desirable, and often the preferred alternative, to do constraint
based checking. We will investigate line of work as another possible
extension to the tool.
\bibliographystyle{plainnat}
\bibliography{paper}
\end{document}